\documentclass[a4paper,11pt]{amsart}

\DeclareMathSizes{11}{11}{7.7}{5.5}
\usepackage{amssymb}
\usepackage{amsmath}
\usepackage[dvips]{graphicx}
\usepackage[top=1.35in,bottom=1.35in,left=1.35in,right=1.35in]{geometry}

\begin{document}
\title {Achievable Transverse Cylindrical Electromagnetic Mode}
\author {R. Chen(1) and X. Li(2) \\}
\address{(1)Shenzhen Institute of Physics and Mathematics, Shenzhen, 518028, China\\}
\address{(2)South China Normal University, Guangzhou, 510631, China\\}

\begin{abstract}
The system of Maxwell equations with an initial condition in a
vacuum is solved in a cylindrical coordinate system. It derives the
cylindrical transverse electromagnetic wave mode in which the
electric field and magnetic field are not in phase. Such
electromagnetic wave can generate and exist in actual application,
and there is no violation of the law of conservation of energy
during the electromagnetic field interchanges.
\end{abstract}

\maketitle


\section{Introduction}
\label{sec:mylabel1}

The general solutions of Maxwell equations \cite{Maxwell:1,Van:1964}
contain unnumbered forms of wave modes in both magnetic field and
electric field, but only the special solutions that obey its initial
or boundary conditions can describe the achievable electromagnetic
wave.

However, classical theories discuss time-harmonic transverse
electromagnetic waves without verifying whether initial conditions
are achieved, but usually predefine the forms of the solutions to
the Maxwell equations
\cite{Jackson:1998,Landau:1980,Tsang:1997,Yang:2006} by adopting the
plural expressions, ${\bf{E}}\left( {{\bf{r}},t} \right)$ = $
{\bf{E}}\left( {\bf{r}} \right){e^{i\omega t}}$ and ${\bf{B}}\left(
{{\bf{r}},t} \right)$ = $ {\bf{B}}\left( {\bf{r}} \right){e^{i\omega
t}}$. Consequently, the electric field and magnetic field of the
time-harmonic electromagnetic wave obtained above are in-phase
everywhere. These electromagnetic modes may form an initial-moment
crisis and violate the conservation of energy during the electric
and magnetic fields interchanges to each other. We note that the
acknowledged results achieved by some previous authors
\cite{Harrington:1961} are actually derived from the above
predefined plural expressions of the solutions of the Maxwell
equations. In classical theory of electromagnetic wave, the
substantial existence of electromagnetic waves is given little
attention to \cite{Brittingham:2009}, because both the propagation
and reception of electromagnetic waves are independent of their
respective phase differences actually. There were no immediate
details about whether one can design and run an experiment to test
if it exists for the electromagnetic wave in which the electronic
field and magnetic field in phase. However, a physical theory must
always guarantee the consistency of its mathematical deductions.
This is just why we restudy the problem of initial value for the
Maxwell equations. According to the theory of differential
equations, solutions of Maxwell equations should not violate natural
initial and boundary conditions \cite{Chen:2000,Chen:2003}. Even if
an excited source and the consequent radiation of the
electromagnetic waves change according to pure oscillations by the
same frequency, the generated electric fields and magnetic fields
cannot be in same phase with each other, otherwise the energy
conservation law would be violated.

We solved the problem of initial value for the plane transverse
electromagnetic mode \cite{Chen:2001}. It shows that the classical
plane electromagnetic wave may be unable to produce practically, but
this conclusion does not have an effect on the general application
of electromagnetic waves
\cite{Coleman:2003,Chu:1982,Davis:1981,Chew:1994,Elliott:1993}.
Unlike the classical plane electromagnetic wave modes, if the time
harmonic wave propagates along the positive direction of x-axis, the
solutions of the Maxwell equations that accord with an initial value
condition in a vacuum would be the following forms \cite{Chen:2001}
\begin{equation}
\label{eq1}
\begin{array}{l}
{\rm {\bf E}} = {\rm {\bf e}}_y E_m \sin \omega \left( {t -
\frac{x}{c}}
\right) \\
{\rm {\bf B}} = {\rm {\bf e}}_z \frac{E_m }{c}\left[ {\sin
\frac{\omega
x}{c} + \sin \omega \left( {t - \frac{x}{c}} \right)} \right] \\
 \end{array}
\end{equation}
\noindent or
\begin{equation}
\label{eq2}
\begin{array}{l}
  {\rm {\bf E}} = {\rm {\bf e}}_y E_m \cos \omega
\left( {t - \frac{x}{c}}\right) \\
  {\rm {\bf B}} = {\rm {\bf e}}_z \frac{E_m }{c}\left[
{\cos \omega \left( {t
- \frac{x}{c}} \right) - \cos \frac{\omega x}{c}} \right] \\
 \end{array}
\end{equation}
\noindent the wave modes obtained above meet with explicit physical
meanings.

Here we follow the work that derives the equations (\ref{eq1}) and
(\ref{eq2}) to discuss the problem of initial value for the
cylindrical transverse electromagnetic mode. We obtain the solutions
of Maxwell equations with a sort of initial condition in a
cylindrical coordinate system. Its electric field and magnetic field
are of axial symmetry, with electric field vectors pointing to the
radical direction of the coaxial cylinder and magnetic field vectors
to the tangential direction. It is similar to the conclusion upon
the plane transverse electromagnetic mode, which is practically able
to produce and exist. That is, the magnetic field and electric field
of the cylindrical transverse electromagnetic wave are in different
phase, and their amplitude ratio depends on the coordinate of space
as well.

\section{Initial conditions of time-harmonic Cylindrical Transverse waves }
\label{sec:mylabel2}

In solving simultaneous differential equations, definite conditions,
including initial conditions and boundary conditions, are required
to be determined \cite{Karlsson:1992}. If the space is infinite,
then we only need to consider the initial condition. In fact, since
the initial moment can be selected at random, the initial condition
of the Maxwell equations does exist in any situation. The coaxial
transmission of electromagnetic waves in hollow conductors and core
wires is presented in Figure 1.1. The radius of the wire is $R_1 $,
and that of the hollow conductor is $R_2 $. In a vacuum, if the wire
and the hollow conductor are having the same amount of positive and
negative charge (the electric quantity per unit length along the
axes on surface is $\lambda _m )$, then the electrostatic field
inside the conductor can be written as
\begin{equation}
\label{eq3}
\mathbf{E}=\frac{{{\lambda }_{m}}}{2\pi {{\varepsilon
}_{0}}r}{{\mathbf{e}}_{r}}\
\end{equation}
\noindent where ${R_1} \le r \le {R_2}$. Among which ${\rm {\bf
e}}_r $ is radical unit vector, and $r$is the distance from a
certain point in the coaxial transmission line to the axis.

\begin{figure}[htbp]
\centerline{\includegraphics[width=3.5088in,height=0.9928in]{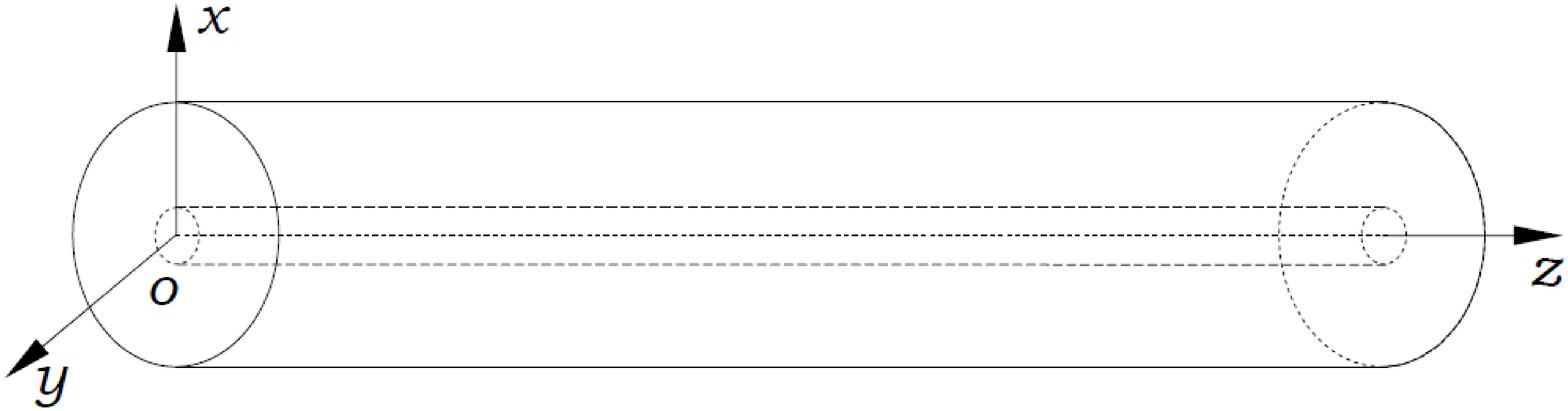}}
\label{fig2}
\end{figure}

\noindent Figure 1. Cylindrical transverse electromagnetic wave
inside the
coaxial transmission line\\

When the left side of the conductor is connected to a periodically
changing power supply, the distribution of charges thus varies and
propagates towards right, which forms electric field waves
transmitting along the positive direction of z, vertical to the
direction of the electric field. If we have the apart of the hollow
conductor with the wire very close, and the electric field steadily
distributed, then, based on the distribution formula of
electrostatic field (\ref{eq3}), the formula of the electric field
wave in a vacuum can also be determined.
\begin{equation}
\label{eq4}
\begin{array}{l}
 {\rm {\bf E}} = {\rm {\bf e}}_r \frac{\lambda _m
}{2\pi \varepsilon _0
r}f\left( {z,t} \right) \\
 E\left( {t = 0} \right) = 0 \\
 E\left( {z = 0} \right) = {\rm {\bf e}}_r
\frac{\lambda _m }{2\pi
\varepsilon _0 r}f\left( t \right) \\
 {\rm {\bf B}}\left( {t = 0} \right) = 0 \\
 \end{array}
\end{equation}
\noindent where $f\left( {z,t} \right)$ may be simple harmonic,
which is determined by the charge distribution on the surface of
conductor. If the excitation sources is an alternating currents of
high frequency, which the oscillation current changes according sine
function, then the corresponding wave mode with the initial and
boundary conditions in a vacuum will be expressed by plural
expressions
\begin{equation}
\label{eq5}
\begin{array}{l}
 {\rm {\bf E}} = {\rm {\bf e}}_r \frac{\lambda _m
}{2\pi \varepsilon _0
r}e^{i{\kern 1pt} \omega \left( {t - \frac{z}{c}} \right)} \\
 E\left( {t = 0} \right) = 0 \\
 E\left( {z = 0} \right) = {\rm {\bf e}}_r
\frac{\lambda _m }{2\pi
\varepsilon _0 r}e^{i{\kern 1pt} \omega t} \\
 {\rm {\bf B}}\left( {t = 0} \right) = 0 \\
 \end{array}
\end{equation}
At the initial moment, the electric field within the transmission
wire is zero, and hence the magnetic field is zero as well, which
build up the initial conditions.

\section{General solutions of Maxwell equations in cylindrical coordinate system}

Periodically changing electric fields generate periodically changing
magnetic fields, and the correspondingly generated electromagnetic
wave propagates at the speed of light in vacuum. The expressions of
the electric field and the magnetic field satisfy Maxwell equations
\begin{equation}
\label{eq6}
\begin{array}{l}
 \nabla \times {\rm {\bf E}} = - \frac{\partial {\rm
{\bf B}}}{\partial t}\\
 \nabla \times {\rm {\bf B}} =
\frac{1}{c^2}\frac{\partial {\rm {\bf
E}}}{\partial t} \\
 \nabla \cdot {\rm {\bf E}} = 0 \\
 \nabla \cdot {\rm {\bf B}} = 0 \\
 \end{array}
\end{equation}
Since the problem is axially symmetric, in cylindrical coordinate
system \cite{Ruan:2007}, the simultaneous equation (\ref{eq6}) can
be written as \cite{Isic:2008}
\begin{equation}
\label{eq7}
\begin{array}{l}
\displaystyle {\rm {\bf e}}_r \left( {\frac{1}{r}\frac{\partial E_z
}{\partial \theta } - \frac{\partial E_\theta }{\partial z}} \right)
+ {\rm {\bf e}}_z \left[ {\frac{1}{r}\frac{\partial }{\partial
\,r}\left( {rE_\theta } \right) - \frac{1}{r}\frac{\partial E_r
}{\partial \theta }} \right]+ {\rm {\bf e}}_\theta \left(
{\frac{\partial E_r }{\partial z} - \frac{\partial E_z }{\partial
\,r}} \right) \\
\displaystyle = - \frac{\partial {\rm {\bf
B}}}{\partial \,t} \\
\displaystyle {\rm {\bf e}}_r \left( {\frac{1}{r}\frac{\partial B_z
}{\partial \theta } - \frac{\partial B_\theta }{\partial z}} \right)
+ {\rm {\bf e}}_z \left[ {\frac{1}{r}\frac{\partial }{\partial
\,r}\left( {rB_\theta } \right) - \frac{1}{r}\frac{\partial B_r
}{\partial \theta }} \right] + {\rm {\bf e}}_\theta \left(
{\frac{\partial B_r }{\partial z} - \frac{\partial B_z }{\partial
\,r}} \right)\\
\displaystyle = \frac{1}{c^2}\frac{\partial
{\rm {\bf E}}}{\partial \,t} \\
\displaystyle \frac{1}{r}\frac{\partial }{\partial \,r}\left( {rE_r
} \right) + \frac{1}{r}\frac{\partial E_\theta }{\partial \theta } +
\frac{\partial E_z}{\partial z} = 0 \\
\displaystyle \frac{1}{r}\frac{\partial }{\partial \,r}\left( {rB_r
} \right) + \frac{1}{r}\frac{\partial B_\theta }{\partial \theta } +
\frac{\partial B_z
}{\partial z} = 0 \\
 \end{array}
\end{equation}
As to electric mode (\ref{eq4}), since ${E_r} = {E_r}\left( {r,z,t}
\right), {E_\theta } = 0 and {E_z} = 0$, simultaneous equations
(\ref{eq7}) can be rewritten as
\begin{equation}
\label{eq8}
\begin{array}{l}
\displaystyle {\rm {\bf e}}_\theta \frac{\partial E_r }{\partial z}
= - {\rm {\bf e}}_r \frac{\partial B_r }{\partial \,t} - {\rm {\bf
e}}_\theta \frac{\partial B_\theta }{\partial \,t} - {\rm {\bf e}}_z
\frac{\partial B_z }{\partial
\,t} \\
\displaystyle {\rm {\bf e}}_r \left( {\frac{1}{r}\frac{\partial B_z
}{\partial \theta } - \frac{\partial B_\theta }{\partial z}} \right)
+ {\rm {\bf e}}_z \left[ {\frac{1}{r}\frac{\partial }{\partial
\,r}\left( {rB_\theta } \right) - \frac{1}{r}\frac{\partial B_r
}{\partial \theta }} \right]  + {\rm {\bf e}}_\theta \left(
{\frac{\partial B_r }{\partial z} - \frac{\partial B_z }{\partial
\,r}} \right) \\= {\rm {\bf e}}_r
\frac{1}{c^2}\frac{\partial E_r }{\partial \,t} \\
\displaystyle \frac{1}{r}\frac{\partial }{\partial \,r}\left( {rE_r } \right) = 0 \\
\displaystyle \frac{1}{r}\frac{\partial }{\partial \,r}\left( {rB_r
} \right) + \frac{1}{r}\frac{\partial B_\theta }{\partial \theta } +
\frac{\partial B_z
}{\partial z} = 0 \\
 \end{array}
\end{equation}
According to the third equation above, the electric field satisfies
the equation $rE_r = g\left( {z,t} \right)$, so that
\begin{equation}
\label{eq9}
E_r = \frac{g\left( {z,t} \right)}{r}
\end{equation}

The result is in accordance with that deduced by (\ref{eq4}), which
means the third equation of (\ref{eq8}) is valid. Because that both
sides of the equation are correspondingly equal to each other, so
that the other three equations can be transformed into
\begin{equation}
\label{eq10}
\begin{array}{l}
\displaystyle \frac{{\partial {B_r}}}{{\partial \,t}} = 0;\;\;\frac{{\partial {B_z}}}{{\partial \,t}} = 0;\;\;\frac{{\partial {B_\theta }}}{{\partial \,t}} =  - \frac{{\partial {E_r}}}{{\partial z}}\quad  \\
\displaystyle \frac{1}{r}\frac{{\partial {B_z}}}{{\partial \theta }} - \frac{{\partial {B_\theta }}}{{\partial z}} = \frac{1}{{{c^2}}}\frac{{\partial {E_r}}}{{\partial \,t}};\;\;\frac{{\partial {B_r}}}{{\partial z}} - \frac{{\partial {B_z}}}{{\partial \,r}} = 0 \\
\displaystyle \frac{1}{r}\frac{\partial }{{\partial \,r}}\left( {r{B_\theta }} \right) - \frac{1}{r}\frac{{\partial {B_r}}}{{\partial \theta }} = 0 \\
\displaystyle \frac{1}{r}\frac{\partial }{{\partial \,r}}\left( {r{B_r}} \right) + \frac{1}{r}\frac{{\partial {B_\theta }}}{{\partial \theta }} + \frac{{\partial {B_z}}}{{\partial z}} = 0 \\
 \end{array}
\end{equation}
According to the first three equations above, we get
\begin{equation}
\label{eq11}
\begin{array}{l}
 B_r = B_r \left( {r,\theta ,z} \right) \\
 B_z = B_z \left( {r,\theta ,z} \right) \\
 B_\theta = - \int {\frac{\partial E_r }{\partial z}dt}
+ a\left( {r,\theta
,z} \right)\quad \\
 \end{array}
\end{equation}
This result will certainly satisfy the last two equations of
(\ref{eq10}). Consequently, the magnetic field can be written as
\begin{equation}
\label{eq12}
{\rm {\bf B}} = {\rm {\bf e}}_r B_r \left( {r,\theta ,z} \right) + {\rm {\bf
e}}_\theta \left[ { - \int {\frac{\partial E_r }{\partial z}dt} + a\left(
{r,\theta ,z} \right)} \right] + {\rm {\bf e}}_z B_z \left( {r,\theta ,z}
\right)
\end{equation}
By using the initial condition ${\rm {\bf B}}\left( {t = 0} \right)
= 0$, we get
\begin{equation}
\label{eq13}
B_r \left( {r,\theta ,z} \right) = 0;\quad B_z \left( {r,\theta ,z} \right)
= 0
\end{equation}
Therefore,
\begin{equation}
\label{eq14}
{\rm {\bf B}} = {\rm {\bf e}}_\theta \left[ { - \int {\frac{\partial E_r
}{\partial z}dt} + a\left( {r,\theta ,z} \right)} \right]
\end{equation}
On the other hand, by the use of (\ref{eq13}), and the integration
of the first equation on the second line of (\ref{eq10}), we get
\begin{equation}
\label{eq15}
{\rm {\bf B}} = {\rm {\bf e}}_\theta \left[ { - \int
{\frac{1}{c^2}\frac{\partial E_r }{\partial \,t}} dz + b\left( {r,\theta ,t}
\right)} \right]
\end{equation}
Based on (\ref{eq4}) and (\ref{eq14}), the general equation of
cylindrical electromagnetic mode can be written as
\begin{equation}
\label{eq16}
\begin{array}{l}
 {\bf{E}} = {{\bf{e}}_r}\frac{{{\lambda _m}}}{{2\pi {\varepsilon _0}r}}f\left( {z,t} \right) \\
 {\bf{B}} = {{\bf{e}}_\theta }\left[ { - \int {\frac{{\partial {E_r}}}{{\partial z}}dt}  + a\left( {r,\theta ,z} \right)} \right] \\
 {\bf{B}}\left( {t = 0} \right) = 0 \\
 \end{array}
\end{equation}
\noindent or
\begin{equation}
\label{eq17}
\begin{array}{l}
 {\bf{E}} = {{\bf{e}}_r}\frac{{{\lambda _m}}}{{2\pi {\varepsilon _0}r}}f\left( {z,t} \right) \\
 {\bf{B}} = {{\bf{e}}_\theta }\left[ { - \int {\frac{1}{{{c^2}}}\frac{{\partial {E_r}}}{{\partial \,t}}} dz + b\left( {r,\theta ,t} \right)} \right] \\
 {\bf{B}}\left( {t = 0} \right) = 0 \\
 \end{array}
\end{equation}
(\ref{eq16}) and (\ref{eq17}) are equivalent to each other. In a
vacuum, they are both general solutions of the Maxwell equations,
where the magnetic field provoked by a radical electric field.

\section{Time-harmonic Cylindrical Transverse Electromagnetic Mode}
\label{sec:mylabel3}

On the basis of (\ref{eq16}) or (\ref{eq17}), the expression of magnetic field can be
acquired by knowing the expression of electric field. By inserting (\ref{eq5}) into
(\ref{eq16}), we get
\begin{equation}
\label{eq18}
\begin{array}{l}
 {\bf{E}} = {{\bf{e}}_r}\frac{{{\lambda _m}}}{{2\pi {\varepsilon _0}r}}{e^{i{\kern 1pt} \omega \left( {t - \frac{z}{c}} \right)}} \\
 {\bf{B}} = {{\bf{e}}_\theta }\left[ { - \int {\frac{{\partial {E_r}}}{{\partial z}}dt}  + a\left( {r,\theta ,z} \right)} \right] \\
 {\bf{B}}\left( {t = 0} \right) = 0 \\
 \end{array}
\end{equation}
According to the first equation, we get
\begin{equation}
\label{eq19}
\frac{\partial E_r }{\partial z} = - \frac{i\omega }{c}\frac{\lambda _m
}{2\pi \varepsilon _0 r}e^{i{\kern 1pt} \omega \left( {t - \frac{z}{c}}
\right)}
\end{equation}
By inserting (\ref{eq19}) into the second equation in (\ref{eq18}),
and then integrating it, we get
\begin{equation}
\label{eq20}
{\bf{B}} = {{\bf{e}}_\theta }\left[
{\frac{1}{c}\frac{{{\lambda _m}}}{{2\pi {\varepsilon
_0}r}}{e^{i{\kern 1pt} \omega \left( {t - \frac{z}{c}} \right)}} +
a\left( {r,\theta ,z} \right)} \right]
\end{equation}
Using the initial conditions of (\ref{eq18}), the undetermined
function can be confirmed
\begin{equation}
\label{eq21}
a\left( {r,\theta ,z} \right) =  -
\frac{1}{c}\frac{{{\lambda _m}}}{{2\pi {\varepsilon
_0}r}}{{\mathop{\rm e}\nolimits} ^{ - i{\kern 1pt} \omega
\frac{z}{c}}}
\end{equation}
After substituting the function into (\ref{eq20}), we further obtain
\begin{equation}
\label{eq22}
{\bf{B}} = {{\bf{e}}_\theta }\frac{{{\lambda
_m}}}{{2\pi c{\varepsilon _0}r}}\left[ {{e^{i{\kern 1pt} \omega
\left( {t - \frac{z}{c}} \right)}} - {e^{ - i{\kern 1pt} \omega
\frac{z}{c}}}} \right]
\end{equation}
Therefore, (\ref{eq18}) can be transformed into
\begin{equation}
\label{eq23}
\begin{array}{l}
 {\bf{E}} = {{\bf{e}}_r}\frac{{{\lambda _m}}}{{2\pi {\varepsilon _0}r}}{e^{i{\kern 1pt} \omega \left( {t - \frac{z}{c}} \right)}} \\
 {\bf{B}} = {{\bf{e}}_\theta }\frac{{{\lambda _m}}}{{2\pi c{\varepsilon _0}r}}\left[ {{e^{i{\kern 1pt} \omega \left( {t - \frac{z}{c}} \right)}} - {e^{ - i{\kern 1pt} \omega \frac{z}{c}}}} \right] \\
 \end{array}
\end{equation}
So we have the expressions of electromagnetic waves (\ref{eq23}) in
a vacuum satisfying every single component equation of Maxwell
(\ref{eq10}), and hence satisfying Maxwell equations. The expression
(\ref{eq23}) is a particular solution of Maxwell's equations that
accord with initial conditions. It represents a kind of existing
transverse cylindrical time-harmonic electromagnetic wave mode. We
obtain the electromagnetic wave mode generated by sinusoidal
electric wave by using the imaginary part of (\ref{eq23}).
\begin{equation}
\label{eq24}
\begin{array}{l}
 {\bf{E}} = {{\bf{e}}_r}\frac{{{\lambda _m}}}{{2\pi {\varepsilon _0}r}}\sin \omega \left( {t - \frac{z}{c}} \right) \\
 {\bf{B}} = {{\bf{e}}_\theta }\frac{{{\lambda _m}}}{{2\pi c{\varepsilon _0}r}}\left[ {\sin \omega \left( {t - \frac{z}{c}} \right) + \sin \frac{{\omega \,z}}{c}} \right] \\
 \end{array}
\end{equation}
By inserting the real part of (\ref{eq23}), the electromagnetic wave
generated by cosine electric wave can be obtained
\begin{equation}
\label{eq25}
\begin{array}{l}
 {\bf{E}} = {{\bf{e}}_r}\frac{{{\lambda _m}}}{{2\pi {\varepsilon _0}r}}\cos \omega \left( {t - \frac{z}{c}} \right) \\
 {\bf{B}} = {{\bf{e}}_\theta }\frac{{\omega {\lambda _m}}}{{2\pi c{\varepsilon _0}r}}\left[ {\cos \omega \left( {t - \frac{z}{c}} \right) - \cos \frac{{\omega \,z}}{c}} \right] \\
 \end{array}
\end{equation}

\section{Conclusions}

From Maxwell' theory, if the electric field in space changes, the
magnetic field at the same place would change and these changing
electric field and magnetic field will generate new changing
electric field and magnetic field in farther space. Consequently,
the changed electric field and magnetic field are not confined to a
region but propagate from near site to farther places. The
propagation of electromagnetic field forms electromagnetic waves.
The characteristics of the electromagnetic wave are actually
described by the solutions  \cite{Naqvi:2010,Chen:2010,Pinel:2010}
of the Maxwell equations. The electromagnetic wave is a transverse
wave, which is just determined by the transverse wave conditions of
the Maxwell equations, $\nabla \cdot {\rm {\bf E}} = 0$ and $\nabla
\cdot {\rm {\bf B}} = 0$.

Since the classical theory first preconcerts the forms of solutions
before solving the Maxwell equations, the obtained transverse
electromagnetic wave mode shows a most incredible characteristic
that the electric field and magnetic field are in phase everywhere.
Now that the periodic changing electric field generate periodic
changing magnetic field, the process of the transmission of an
electromagnetic wave is that of energy radiation, and the
electromagnetic energy is transmitted with the electromagnetic
waves. If the maximum or minimum value of electric field and
magnetic field in an electromagnetic wave are always obtained
simultaneously, how do the electric energy and magnetic energy
interchange? Consequently, we conclude that the classical
electromagnetic wave in which the electric field and magnetic field
are in phase everywhere violates the law of conservation of energy.

We strive to maintain the conclusions of classical theory, but the
initial condition of Maxwell equations has always been there. We
wish to know the solutions of the Maxwell equations with a certain
initial condition. The cylindrical transverse electromagnetic wave
modes in a vacuum are hence derived. It can be determined that, with
the solutions of Maxwell equations obtained under initial values,
the existing electric field vector and magnetic field vector, the
velocity vector of the plane or cylindrical electromagnetic wave are
vertical to one another. Nevertheless, the amplitude ratio of
electric field and magnetic field is not constant in space, and
their phase difference changes as well, so that they do not reach
maximum or zero simultaneously.

\end{document}